\documentclass{article}  
\usepackage{graphicx} 
\usepackage{amsmath,amssymb,amsbsy,amstext, amsthm, simplewick}
\usepackage{xcolor}
\usepackage{mdframed}
\mdfsetup{%
linecolor=white,
backgroundcolor=gray!20,
}
\usepackage{amsfonts}
\usepackage{jcappub}
\def\be{\begin{equation}}
\def\ee{\end{equation}}
\def\bea{\begin{eqnarray}}
\def\eea{\end{eqnarray}}
\def\com{{\rm com}}
\newcommand{\llp}{\left [}
\newcommand{\rrp}{\right ]}
\newcommand{\lp}{\left (}
\newcommand{\rp}{\right )}
\def\PBH{{\rm PBH}}
\def\lsim{\mathrel{\rlap{\lower4pt\hbox{\hskip0.5pt$\sim$}}
    \raise1pt\hbox{$<$}}}         
\def\gsim{\mathrel{\rlap{\lower4pt\hbox{\hskip0.5pt$\sim$}}
    \raise1pt\hbox{$>$}}}         

\makeatletter
\newlength{\apb@width}
\newcommand{\autoparbox}[2][c]{\settowidth{\apb@width}{#2}\parbox[#1]{\apb@width}{#2}}
\newcommand{\includegraphicsbox}[2][]{\autoparbox{\includegraphics[#1]{#2}}}

\begin{document}

\begin{titlepage}

\setcounter{page}{1} \baselineskip=15.5pt \thispagestyle{empty}

\bigskip\

\vspace{2cm}
\begin{center}

{\fontsize{20}{28}\bfseries Renormalized Primordial Black Holes}

\end{center}

\vspace{0.5cm}

\begin{center}
{\fontsize{13}{30}\selectfont  G. Franciolini,$^{1}$ A. Ianniccari,$^{2}$ A. Kehagias, $^{2,3}$ 
D. Perrone,$^{2}$  and A. Riotto,$^{2}$}
\end{center}

\begin{center}

\vskip 8pt
\textsl{$^1$ CERN, Theoretical Physics Department,
Esplanade des Particules 1, Geneva 1211, Switzerland}
\vskip 7pt

\textsl{$^2$ Department of Theoretical Physics and Gravitational Wave Science Center (GWSC), \\
 Université de Genève, CH-1211 Geneva }
\vskip 7pt

\textsl{$^3$ Physics Division, National Technical University of Athens, Athens, 15780, Greece}
\vskip 7pt

\end{center}

\vspace{1.2cm}
\hrule \vspace{0.3cm}
{ \noindent {\fontsize{12}{12} \bfseries Abstract} \\[0.1cm]
The formation of primordial black holes in the early universe may happen through the collapse of large curvature perturbations generated during a non-attractor phase of inflation or through a curvaton-like dynamics after inflation. The fact that such small-scale  curvature perturbation is 
 typically non-Gaussian  leads to the renormalization of composite operators built up from  the smoothed density contrast and  entering in the calculation   of the primordial black abundance. Such renormalization causes  the phenomenon of operator mixing and  the appearance of an infinite tower of local, non-local and  higher-derivative operators as well as to a sizable shift in the threshold for primordial black hole formation. 
 This hints that the calculation of the primordial black hole abundance is more involved than what generally assumed.

\noindent}
\vspace{0.2cm}
\hrule

\vspace{0.6cm}

\end{titlepage}

 \tableofcontents

\newpage 
\baselineskip=18pt
\section{Introduction}
The  
physics of Primordial Black Holes (PBHs)~\cite{Sasaki:2018dmp, Carr:2020gox, Green:2020jor,LISACosmologyWorkingGroup:2023njw} has recently attracted much attention thanks to the plethora of gravitational wave detections from the mergers of BH binaries~\cite{LIGOScientific:2016aoc, LIGOScientific:2018mvr,LIGOScientific:2020ibl, LIGOScientific:2021djp}. Maybe even more interesting, some of the LIGO/Virgo/KAGRA events might be in fact  originating from the mergers of PBHs \cite{Bird:2016dcv,Sasaki:2016jop,Clesse:2016vqa,Ali-Haimoud:2017rtz,Hutsi:2020sol, DeLuca:2021wjr,Franciolini:2021tla,Franciolini:2022tfm}.
Future   gravitational wave  experiments might help in  shedding light on the possible existence of PBHs \cite{Chen:2019irf,Pujolas:2021yaw,DeLuca:2021hde,Barsanti:2021ydd,Bavera:2021wmw}.

In this paper we will start from the rather standard point of view that  PBHs  are  born in the radiation-dominated era   by the collapse of large overdensities generated on small scales during a non-attractor phase of inflation or through a curvaton-like dynamics after inflation ~\cite{Sasaki:2018dmp}. 

Calculating the abundance of PBHs as precisely as possible is a crucial step in assessing if PBHs may account for a substantial fraction of the dark matter of the universe or if they satisfy the different observational constraints. Such computation is rendered difficult as PBHs are rare events and therefore the formation probability is sensitive to small changes in the various ingredients, such as  the critical threshold of collapse \cite{Musco:2018rwt,Musco:2020jjb}, the non-Gaussian nature of the fluctuations \cite{DeLuca:2019qsy,Young:2019yug}, the choice of the window function to define smoothed observables \cite{Young:2019osy}, to mention a few. Furthermore it has been recently pointed out that non-linear corrections entering in the calculation of the PBHs abundance from  the 
 non-linear radiation transfer function and 
the determination of the true physical horizon crossing  corrects
 the overdensity (on  comoving orthogonal slices) by introducing many   
 non-linear terms, even of non-local type \cite{DeLuca:2023tun}. This makes the calculation of the formation
probability highly nontrivial.

In this paper we point out that the standard way of calculating the PBH abundance, which we will revise in the next section, overlooks a technical, but simple point. In the  standard calculation one defines a  smoothed density contrast $\delta_m(r_m,\vec x_{\rm pk})$, 
which depends on the position $\vec x_{\rm pk}$ of the peak of the profile of the overdensity and on the distance $r_m$ from it where the compaction function has its maximum. This field 
turns out to be the sum of a field $\delta(r_m,\vec x_{\rm pk})$,  but also its square $\delta^2(r_m,\vec x_{\rm pk})$. The latter is a  composite operator, i.e., evaluated at the same point of space (and time), and receives loop contributions from all scales. This happens because $\delta(r_m,\vec x_{\rm pk})$ is proportional to the gradient of the small-scale curvature perturbation and  the latter is typically non-Gaussian (see e.g., \cite{Ferrante:2022mui}).  In other words the curvature perturbation has cubic and quartic interactions, leading to the renormalization of the composite operator $\delta^2(r_m,\vec x_{\rm pk})$ and therefore of $\delta_m(r_m,\vec x_{\rm pk})$.

The renormalization procedure leads  to the well-known phenomenon of operator mixing: the renormalized operator $\delta^2(r_m,\vec x_{\rm pk})$ receives corrections from all possible operators compatible with the symmetries of the problem. We will show that indeed one-loop corrections assume the form of an infinite tower of operators, even non-local in $\delta(r_m,\vec x_{\rm pk})$ and of the higher-derivative type. It is the interacting nature of the  comoving curvature perturbation which gives rise to the phenomenon of operator mixing, which leads to a shift in the correlators and the threshold for PBH formation, as well as a modification of the formation probability.

The paper is organized as follows. In section 2 we  set the stage and describe briefly the standard way of calculating the PBH abundance. In section 3 we remind the reader why PBH formation models are characterized by a non-Gaussian curvature perturbation. In section 
4 we introduce the concept of renormalization of the PBHs and the Feynman rules we will adopt to calculate the operator mixing at one-loop, which we do in sections 5 and 6. Section 7 contains our conclusions, while we devote the appendices A and B to the technical details.

\section{ A brief summary of the  standard PBH abundance calculation}
The goal of this  section is to set the stage and  briefly recap  what is the standard procedure to calculate the PBH abundance in the literature, see for instance Refs.~\cite{Young:2019yug, Biagetti:2021eep}. 

As we already wrote,  our focus will be the  PBH formation from the collapse of large  overdensities which are  generated during inflation and re-enter the Hubble radius  in the  radiation-dominated era. The key starting  object  is the  curvature perturbation $\zeta$ appearing  in the metric in the comoving uniform-energy density gauge 
\be
{\rm d}s^2=-{\rm d}t^2+a^2(t)e^{2\zeta(\vec x)}{\rm d}{\vec x}^2.
\ee
Here $a(t)$ is the scale factor in terms of the cosmic time.
Applying  the gradient expansion on superhoizon scales ~\cite{Shibata:1999zs}, one can   relate the non-linear density contrast $\delta_\com(\vec x,t)$  on  comoving orthogonal slicings and the time independent curvature perturbation $\zeta(\vec x)$ ~\cite{Harada:2015yda}
\be
\label{deltaNL}
\delta_\com(\vec x,t)=-\frac{8}{9}\frac{1}{a^2H^2} e^{-5\zeta(\vec x)/2}\nabla^2 e^{\zeta(\vec x)/2},
\ee
where $H$ is the Hubble rate.
Cosmological perturbations may  gravitationally collapse to  form a PBH depending upon   the amplitude measured at the peak of the compaction function, defined to be the mass excess compared to the background value in a given radius~\cite{Harada:2015yda, Musco:2018rwt,Escriva:2019phb}. On superhorizon scales  and assuming spherical symmetry,  it reads
\be
\mathcal{C} (r) = - \frac{2}{3} r \partial_r\zeta (r) \llp 2 +  r \partial_r\zeta (r) \rrp.
\ee
The compaction function has a maximum at the comoving length scale $r_m$ 
such that

\be
\label{max}
\left.\partial_r\zeta(r)+r_m\partial_r^2\zeta(r)\right|_{r=r_m}=0,
\ee
and 
one can  define  a smoothed perturbation amplitude as the volume average of the energy density contrast within the scale $r_m$ at the  Hubble radius crossing time $t_H$~\cite{Musco:2018rwt}
\be
\label{sh}
\delta_m=\frac{3}{\left(r_m e^{\zeta(r_m)}\right)^3}\int_0^{r_m} {\rm d} r\,\delta_\com(r,t_H)\left(r e^{\zeta(r)}\right)^2\partial_r\left(r e^{\zeta(r)}\right),
\ee
where a top-hat window function needs to be used  for a  proper treatment of the threshold~\cite{Young:2019osy}.

The quantity $\delta_m$ is crucial in determining the abundance of PBHs. If  computed at the Hubble radius  crossing time $t_H$, that is 
\be
\epsilon(t_H)=\frac{r_H}{r_m e^{\zeta(r_m)}}=\frac{1}{r_m e^{\zeta(r_m)} \, a H }=1,
\ee
it  becomes
\be
\label{deltam}
\delta_m = \delta_l - \frac{3}{8} \delta_l^2, \qquad \delta_l = - \frac{4}{3}r_m\left.\partial_r\zeta (r)\right|_{r=r_m}.
\ee
The PBH abundance is subsequently   calculated  integrating the probability distribution function of the smoothed density contrast from a threshold value $\delta_c$ on
\be
\beta = \int_{\delta_c} {\rm d} \delta_m\lp \frac{M_\PBH}{M_H} \rp P(\delta_m),
\ee
as a function of the PBH mass $M_\PBH$ and the mass $M_H$ enclosed in the Hubble volume.  
One  can then use the conservation of probability to write~\cite{Germani:2019zez,Young:2019yug,Biagetti:2021eep,DeLuca:2022rfz,Ferrante:2022mui,Gow:2022jfb}
\be
P(\delta_l) {\rm d} \delta_l = P(\delta_m) {\rm d} \delta_m,
\ee
 with threshold given by~\cite{Young:2019yug}
\be
\label{threhsold_delta}
\delta_{l,c} = \frac{4}{3} \lp 1- \sqrt{1-\frac{3}{2}\delta_{c}} \rp,
\ee
where $\delta_{c}$ is the threshold routinely computed by running numerical simulations \cite{Musco:2018rwt}.

For Gaussian curvature perturbations, the probability of the linear density contrast is Gaussian and is exponentially sensitive to the threshold $\delta_{l,c}$ and the variance $\sigma^2_l$
\begin{align}
\label{sigma2}
P(\delta_l)&=\frac{1}{\sqrt{2\pi}\sigma_l} e^{-\delta_{l}^2/2\sigma_l^2},\nonumber\\
\sigma_l^2&= \int\frac{{\rm d}k}{k} {\cal P}_{\delta_l}(k),
\end{align}
where ${\cal P}_{\delta_l}$ is the dimensionless power spectrum of the density contrast $\delta_l$.

\section{The non-Gaussianity of the curvature perturbation}
The probability (\ref{sigma2}) is valid only if the smoothed density contrast $\delta_l$ is linear (hence the ${}_l$ subscript). Given the relation 

\begin{equation}
   \delta_l(r_m) = - \frac{4}{3}r_m \left.\partial_r\zeta (r)\right|_{r=r_m},
\end{equation}
one concludes that $\delta_l$ is indeed linear only if the curvature perturbation is linear. 
However, in all models in which a large fluctuation in the curvature perturbation is generated, the curvature perturbation is  typically non-Gaussian (see,  for instance,  Ref. \cite{Ferrante:2022mui})\footnote{We remark that the non-Gaussianity we are discussing here is induced at small scales and therefore is not   the non-Gaussianity of the curvature perturbation on CMB scales \cite{Bartolo:2004if}, which    are much larger than those interested in the PBH formation. On CMB scales the  non-Gaussianity  is  severely constrained by CMB data \cite{Planck:2019kim}.}. Non-Gaussianity among the modes interested in the growth of the curvature perturbation is generated either by their self-interaction during the ultra slow-roll phase \cite{Cai:2018dkf} or after Hubble radius exit when the curvature perturbation is sourced by a curvaton-like field \cite{Bartolo:2003jx,Sasaki:2006kq}. 

Even though in general the exact  relation between $\zeta$ and its corresponding Gaussian component $\zeta_g$ can be worked out model by model, we will adopt  a cubic  expansion to be model independent,  

\begin{equation}
\label{eq:zeta_to_zetag}
\zeta=\zeta_g+ \frac{3}{5}f_{NL}\zeta_g^2+   \frac{9}{25}g_{NL}\zeta_g^3,
\end{equation}
which is a good approximation if  $(3/5)f_{NL}\zeta_g$ and $(9/25)g_{NL}\zeta^2_g\lsim 1$. For instance, in models with a sharp transition between the ultra slow-roll phase and the subsequent slow-roll phase one has $f_{NL}\simeq 5/2$ and $g_{NL}\simeq 25/6$ (see, for instance, Ref. \cite{Cai:2018dkf}) and therefore the expansion is motivated as long as $\zeta_g\lsim 0.7$. The expansion is also justified if we  perform a perturbative loop expansion and think the  non-Gaussianity as leading to three-, four-point  vertices, and so on. 

The non-Gaussianity of the curvature perturbation is crucial in changing the PBH abundance. Indeed, 
around peaks of the power spectrum of the curvature perturbation,  a positive $f_{NL}$ increases the abundance of the PBHs, while a negative $f_{NL}$ has the opposite effect, thus helping to interpret the  recent pulsar timing array observation  of a stochastic gravitational wave background  \cite{NG15-SGWB,NG15-pulsars,EPTA2-SGWB,EPTA2-pulsars,EPTA2-SMBHB-NP,PPTA3-SGWB,PPTA3-pulsars,PPTA3-SMBHB,CPTA-SGWB} by using the scalar-induced gravitational waves  sourced along with PBH formation~\cite{Franciolini:2023pbf,Liu:2023ymk,Wang:2023ost,Cai:2023dls,Inomata:2023zup,Figueroa:2023zhu,Yi:2023mbm,Zhu:2023faa,Ellis:2023oxs}. 

The importance of the non-Gaussianity in the curvature perturbation will become clear in the following section, where we will go back to the smoothed density contrast  $\delta_m$ and deal with the fact that it is a composite operator.

\section{Renormalization in the PBHs}
We start again from the expression (\ref{deltam}) of the smoothed density contrast

\begin{equation}
    \delta_m(r_m,\vec x_{\rm pk}) = \frac{3}{4\pi r_m^3}\int{\rm d}^3x \,
    \delta_{\rm com}(\vec x,t_H)\,\theta(r_m-|\vec x-\vec x_{\rm pk}|)
    =\delta(r_m,\vec x_{\rm pk}) - \frac 38 \delta^2(r_m,\vec x_{\rm pk}),
\end{equation}
where this time we removed the subscript ${}_l$ to account for the fact that $\delta_l$ is not a Gaussian field, since the curvature perturbation is not. Furthermore we have highlighted the fact that $\delta_m$ depends on the peak position $\vec x_{\rm pk}$ and from the distance $r_m$ from the peak. 

\begin{figure}[t]
	\centering
\includegraphics[width=1\textwidth]{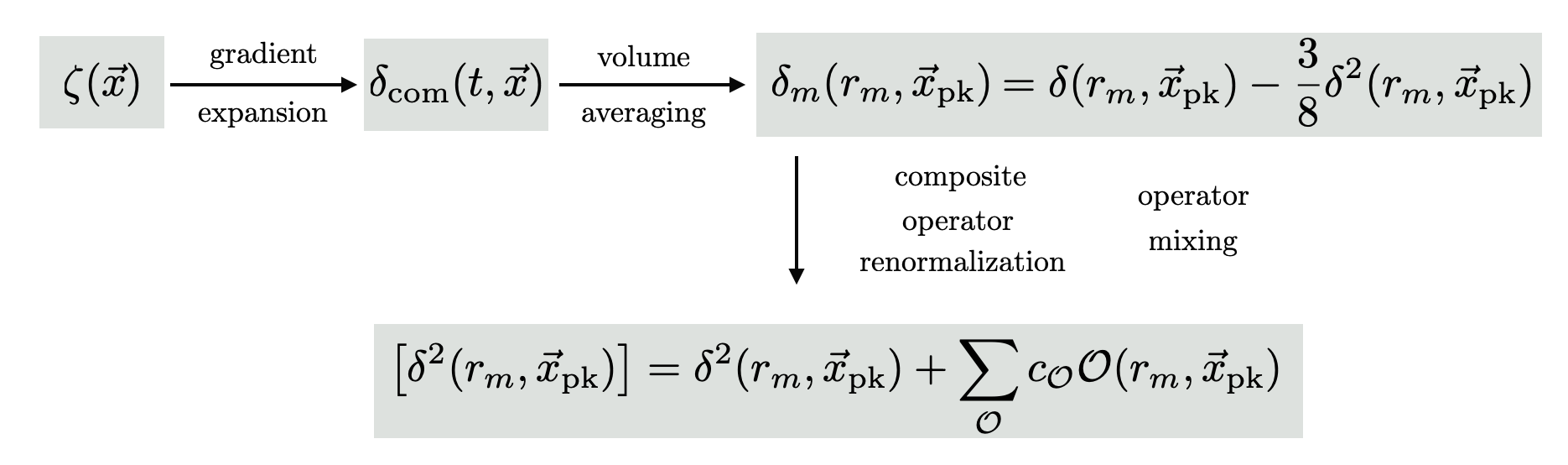}
	\caption{Schematic view of operations the leading to the composite operator renormalization.}
	\label{schema}
\end{figure}
The key point is that $\delta_m(r_m,\vec x_{\rm pk})$ contains the  composite operator $\delta^2(r_m,\vec x_{\rm pk})$ which  receives contributions from all scales, not only from the infrared ones. This is because  two small-scale modes may combine in the loops to form a long mode. Such contributions may be therefore not small at all,  even on large scales. One should also recall at this stage that, even in the presence of the window function cutting-off ultraviolet momenta, variances are typically dominated by integrating momenta corresponding to  length scales  well beneath    the horizon \cite{DeLuca:2023tun}.

One needs to renormalize the composite operator $\delta^2(r_m,\vec x_{\rm pk})$ at the point $\vec x_{\rm pk}$ in such a way that
all potentially large contributions can be systematically removed
by adding local counterterms (canceling as well the dependence on an unphysical  cut-off scale, if needed)
\be
\left[\delta^2(r_m,\vec x_{\rm pk})\right]\equiv \delta^2(r_m,\vec x_{\rm pk})+\sum_{\cal O}c_{ \cal O}{\cal O}(r_m,\vec x_{\rm pk}),
\ee
where the square brackets indicate the renormalized operator. This means that the renormalized composite operator will mix with possibly all the other operators ${\cal O}(r_m,\vec x_{\rm pk})$ which are allowed by the symmetries of the problem. It will also have an impact on the PBH abundance, being it so sensitive to minute changes in the
threshold  and/or in the variances.

The goal of the subsequent sections is to perform the systematic renormalization of the smoothed density contrast.

\subsection{Feynman rules}
Our starting point is the expression
\begin{equation}
\label{start}
     \delta(r_m,\vec x_{\rm pk}) = -\frac{4}{3} r \partial_r \zeta(r_m,\vec x_{\rm pk}) =
    -\frac 49 \int \frac{{\rm d}^3 k}{(2\pi)^3}\; e^{i \vec k\cdot \vec x_{\rm pk}} k^2r_m^2 W_3(k r_m)\zeta_{\vec k} ,
\end{equation}
whose Fourier transform reads
\begin{equation}
    \delta_{\vec k}(r_m) = -\frac 49   k^2r_m^2  W_3(k r_m) \zeta_{\vec k}.
\end{equation}
Here $W_3(k r_m$) is the Fourier transform of the top-hat window function in real space with radius $r_m$
\begin{equation}
    W_3(x)= 3 \frac{\sin x - x \cos x}{x^3}=2^{3/2}\Gamma(5/2)\frac{J_{3/2}(x)}{x^{3/2}},
\end{equation}
being $J_\nu(x)$ the Bessel function of the first kind. To  simplify the  notation we have absorbed the time-dependent radiation transfer function  in the curvature perturbation $\zeta_{\vec k}$ (and  in  its  power spectrum).  All  power spectra will be intended to be calculated at horizon crossing.

The renormalization of the composite operator $\delta^2(r_m,\vec x_{\rm pk})$ will proceed through the following Feynman rules:
\begin{itemize}
    \item draw every $N$-point connected graph and conserve momentum in each vertex;
    
    \item to each $(n+1)$ point vertex, with a solid line (in-going momentum) and $n$ dashed lines, it  corresponds the following vertex
    \begin{align}
        \includegraphicsbox[scale=0.9]{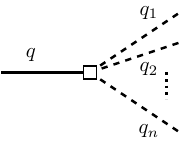} \ &=(-1)^{n-1} n!\, F_n\;  \;\frac{(qr_m)^2 W_3(qr_m) }{\left(\frac{4}{9}\right)^{n-1} \prod_{i=1}^n (q_i r_m)^2 W_3(q_i r_m)},
    \end{align}
    where in the numerator there is the factor $(4/9)(qr_m)^2 W_3(qr_m)$ for the solid line and at the denominator a product of all the factors $(4/9)(q_i r_m)^2 W_3(q_i r_m)$ for each  dashed lines;
    \item the propagator, represented with a dashed line,  corresponds to 
    \begin{equation}
        \includegraphicsbox[scale=0.9]{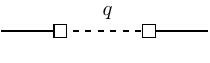} \ = P_{\delta}(q)=\frac{16}{81} r_m^4q^4\,W^2_3(q r_m)P_{\zeta_g}(q);
    \end{equation}
    
    \item integrate for every loop 
    \begin{equation}
         \int_{\vec q}=\int \frac{d^3 q}{(2\pi)^3};
    \end{equation}
    \item multiply a Dirac delta with all the external momenta, for the global momentum conservation
    \begin{equation}
        (2\pi)^3 \delta^{(3)}\left( \sum_{i=1}^N \vec k_i\right);
    \end{equation}
\end{itemize}
 Here $F_n$ is a function that depends on the number of legs in the vertex, for example from Eq. (\ref{eq:zeta_to_zetag})
    \begin{equation}
        F_1 = 1, \quad F_2 = \frac 35 f_{NL}, \quad F_3 = \frac {9}{25}g_{NL}.
    \end{equation}
Notice that for the $n=1$ vertex we have 
\begin{equation}
    \includegraphicsbox[scale=0.9]{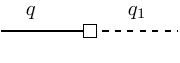} = \frac{(qr_m)^2 W_3(qr_m) }{(q_1r_m)^2 W_3(q_1 r_m)} = 1,
\end{equation}
which is just a consequence of momentum conservation.

\section{Renormalization at one-loop: linear operator mixing}
By indicating the 
 composite operator with the following 
 vertex
 
\begin{equation}
    \delta^2=\includegraphicsbox[scale=0.8]{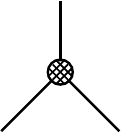}
\end{equation}
its   expectation value  is obtained from the Feynman diagram

\begin{equation}
    \includegraphicsbox[scale=0.8]{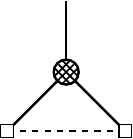} \ = (2\pi)^3 \delta^{(3)}( \vec k) \int_{\vec q} P_{\delta}(q),
\end{equation}
which gives
\begin{equation}
\langle \left(\delta^2\right)_{\vec k}\rangle'=\int_{\vec q} P_\delta(q)=
\frac{16}{81}r_m^4\int_{\vec q} q^4\,W^2_3(qr_m)P_{\zeta_g}(q)=\sigma^2_{\delta(r_m)}. 
\end{equation}
Here and in the following  primes indicate that we remove the factors of $(2\pi)^3$ times the Dirac delta for the momentum conservation. This vacuum expectation value  is removed by adding a constant counterterm

\begin{equation}
\left[\delta^2(r_m,\vec x_{\rm pk})\right]=\delta^2(r_m,\vec x_{\rm pk})-\sigma^2_{\delta(r_m)}.    
\end{equation}
Subtracting this tadpole contribution ensures that the vacuum expectation value of $\delta_m$  vanishes  at the loop level. Furthermore, in the following calculations this constant counterterm will only
contribute to disconnected graphs.\\

The one-loop contribution to $\langle(\delta^2)_{\vec k_1}\delta_{\vec k_2}\rangle$ is,
diagramatically\footnote{We will ignore from now on diagrams such as  
\vskip -0.1cm
\begin{equation}
\includegraphicsbox[scale=0.7]{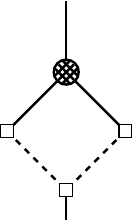}\nonumber
\end{equation}
\vskip 0.1cm
which does not have power spectra of external momenta and lead to contact terms, with no new operators.}

\begin{equation}
\langle(\delta^2)_{\vec k_1}\delta_{\vec k_2}\rangle'=2\times
\includegraphicsbox[scale=0.8]{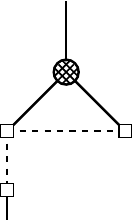}\;,
\end{equation}
which gives

\begin{eqnarray}
\label{eq:threept}
 \langle(\delta^2)_{\vec k_1}\delta_{\vec k_2}\rangle'   &=&-\frac{12}{5}f_{NL} \int_{\vec q}\; \frac{(|\vec k_1-\vec q|r_m)^2 W_3(|\vec k_1-\vec q|r_m)}{\frac{4}{9} (k_2 r_m)^2 W_3(k_2 r_m)(qr_m)^2 W_3(qr_m)} \;
     P_{\delta}(k_2) P_{\delta}(q)\nonumber\\
     &=&-\frac{12}{5}\cdot
     \frac 49\cdot  \frac {16}{81} f_{NL}\int_{\vec q} 
    r_m^6 q^2   |\vec k_1-\vec q|^2 k_2^2
    \; W_3\left(|\vec k_1-\vec q|r_m \right)W_3(k_2 r_m) W_3(qr_m)P_{\zeta_g}(q)P_{\zeta_g}(k_2).\nonumber\\
    &&
\end{eqnarray}
We now use Eq. (\ref{eq:j_splitting}) to perform the angular integral and obtain

\begin{eqnarray}
\langle(\delta^2)_{\vec k_1}\delta_{\vec k_2}\rangle'   &=&
I^{(1)}_{\rm one-leg}+I^{(2)}_{\rm one-leg},\nonumber\\
I^{(1)}_{\rm one-leg}&=&
-\frac{256}{1215}f_{NL}\int_{\vec q} 
    r_m^6 q^2   (q^2+k_2^2) k_2^2
    \; W_3^2(k_2 r_m) W^2_3(qr_m)P_{\zeta_g}(q)P_{\zeta_g}(k_2)\nonumber\\
    I^{(2)}_{\rm one-leg}&=&-\frac{256}{1215} f_{NL}\int_{\vec q} 
    r_m^6 q^2 k_2^2
    \; W^2_3(k_2 r_m) W^2_3(qr_m)\left(  \frac{q^2 }{3}\frac{{\rm d}\ln W_3(k_2 r_m)}{{\rm d} \ln k_2 r_m}  + \; \frac{k_2^2}{3}\frac{{\rm d}\ln W_3(qr_m) }{{\rm d}\ln  qr_m} \right)\nonumber\\
    &\cdot&P_{\zeta_g}(q)P_{\zeta_g}(k_2)\;.\nonumber\\
    &&
    \end{eqnarray}
The first integral gives

\begin{equation}
    I^{(1)}_{\rm one-leg}=\frac{12}{5}f_{NL}\left(\sigma^2_{\delta(r_m)}\langle\zeta_{g\vec k_1}(r_m)\delta_{\vec k_2}(r_m)\rangle'+\frac{4}{9}\sigma^2_{\nabla\zeta_g(r_m)}\langle\delta_{\vec k_1}(r_m)\delta_{\vec k_2}(r_m)\rangle'\right),
\end{equation}
where 

\begin{equation}
\label{eq:zeta_nonlocal_delta}
    \zeta_g(r_m,\vec x_{\rm pk})=
    \int\frac{{\rm d}^3 k}{(2\pi)^3}\,
    e^{i\vec k\cdot\vec x_{\rm pk}} \zeta_{g\vec k}(r_m)=
\int\frac{{\rm d}^3 k}{(2\pi)^3}\,
    e^{i\vec k\cdot\vec x_{\rm pk}}W_3(k r_m) \zeta_{g\vec k} 
\end{equation}
and 
\begin{equation}
    \sigma^2_{\nabla\zeta_g(r_m)} = \int_{\vec q} r_m^2q^2 W_3^2(qr_m)  P_{\zeta_g}(q).
\end{equation}
The  second integral gives

\begin{eqnarray}
    I^{(2)}_{\rm one-leg}=\frac{12}{5}f_{NL}\left[\frac{1}{3}\sigma^2_{\delta(r_m)} \frac{{\rm d}}{{\rm d}\ln r_m}\langle\zeta_{g\vec k_1}(r_m)\delta_{\vec k_2}(r_m)\rangle'+\frac{4}{9}\cdot\left(\frac{1}{6}\frac{{\rm d} \sigma^2_{\nabla\zeta_g(r_m)}}{{\rm d}\ln r_m}-\frac{1}{3}\sigma^2_{\nabla\zeta_g(r_m)}\right)\langle\delta_{\vec k_1}\delta_{\vec k_2}\rangle'\right].\nonumber\\
    &&
\end{eqnarray}
Therefore, the renormalized operator at the level of linear mixing becomes (see Appendix \ref{AppendixB})

\begin{mdframed}
\begin{eqnarray}
\left[\delta^2(r_m,\vec x_{\rm pk})\right]_{\text{linear }}&=&\delta^2(r_m,\vec x_{\rm pk})-\sigma^2_{\delta(r_m)}-\frac{12}{5}f_{NL}\cdot \frac{4}{9}\left(\frac{2}{3}\sigma^2_{\nabla\zeta_g(r_m)}+\frac{1}{6}\frac{{\rm d}\sigma^2_{\nabla\zeta_g(r_m)}}{{\rm d}\ln r_m}\right)\delta(r_m,\vec x_{\rm pk})\nonumber\\
&+&\frac{12}{5}f_{NL}\cdot\frac{5}{3}\cdot\sigma^2_{\delta(r_m)}
\zeta(r_m,\vec x_{\rm pk}),\nonumber
\end{eqnarray}
\end{mdframed}
\begin{eqnarray}
\label{firstcomposite}
&=&\hskip 0.5cm \includegraphicsbox[scale=0.8]{Figures/D_loop_1leg_t.pdf}\hskip 0.5cm
+\hskip 0.5cm\includegraphicsbox[scale=1]{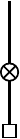}
\end{eqnarray}
where we have added the corresponding Feynman diagram for the counterterm. We see that already at this level there appears a non-local operator in the density smoothed contrast  field  $\delta(r_m,\vec x_{\rm pk})$ once the following identity 

\begin{equation}
    \zeta_g(r_m,\vec x_{\rm pk})=\frac{9}{4r_m^2\nabla^2_{x_{\rm pk}}}\delta(r_m,\vec x_{\rm pk})
\end{equation}
is adopted.

\section{Renormalization at one-loop: quadratic operator mixing}
The one-loop contribution to $\langle[\delta^2]_{\vec k_1}\delta_{\vec k_2}\delta_{\vec k_3}\rangle$ is
diagramatically

\begin{equation}
\label{two}
\langle[\delta^2]_{\vec k_1}\delta_{\vec k_2}\delta_{\vec k_3}\rangle'\hskip 0.3cm =\hskip 0.3cm
\includegraphicsbox[scale=0.8]{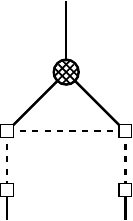}\hskip 0.3cm+\hskip 0.3cm
2\times
\includegraphicsbox[scale=0.8]{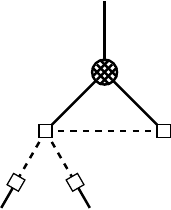}\hskip 0.3cm+\hskip 0.3cm\includegraphicsbox[scale=0.85]{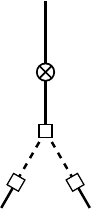} 
\end{equation}
Let us start with the first diagram, which gives

\begin{eqnarray}
    I^{(1)}_{\rm two-legs}&=& \frac{36}{25} f^2_{NL} P_{\delta}(k_2) P_{\delta}(k_3) \left(\frac{9}{4}\right)^2\int_{\vec q} P_{\delta}(q) \frac{(|\vec k_3 + \vec q|r_m)^2  W_3(|\vec k_3 + \vec q|r_m)}{(qr_m)^2 W_3(qr_m) (k_3r_m)^2 W_3(k_3 r_m)}  \nonumber \\
    &\cdot&\frac{(|\vec k_2 - \vec q|r_m)^2 \; W_3(|\vec k_2 - \vec q|r_m)}{(qr_m)^2 W_3(qr_m) (k_2r_m)^2 W_3(k_2 r_m)}.
\end{eqnarray}
Expanding the power spectra we get

\begin{eqnarray}
\label{I3}
    I^{(1)}_{\rm two-legs}&=& \frac{36}{25} f^2_{NL} \left(\frac{16}{81}\right)^2 (k_2r_m)^2 W_3(k_2 r_m) P_{\zeta_g}(k_2) (k_3r_m)^2 W_3(k_3 r_m) P_{\zeta_g}(k_3)  \nonumber \\
    &\cdot& \int_{\vec q} P_{\zeta_g}(q)\; (|\vec k_3 + \vec q|r_m)^2 \; W_3(|\vec k_3 + \vec q|r_m)(|\vec k_2 - \vec q|r_m)^2 \; W_3(|\vec k_2 - \vec q|r_m).
\end{eqnarray}
The calculation is rather involved and we do not report it here in full length, the interested reader can find all the details in the Appendix \ref{appendix:2window_fun}. The procedure once again is to disentangle the sum of the vectors in the window functions using the addition theorem of the Bessel functions. The integral (\ref{I3}) is then written as

\begin{eqnarray}
    I^{(1)}_{\rm two-legs}&=& \frac{36}{25} f^2_{NL} \left(\frac{4}{9}\right)^4 (k_2r_m)^2 W_3(k_2 r_m) P_{\zeta_g}(k_2) (k_3r_m)^2 W_3(k_3 r_m) P_{\zeta_g}(k_3)\nonumber\\
    &\cdot&r_m^4\int_{\vec q} P_{\zeta_g}(q)\;\left[k_2^2 k_3^2 I^{00}(\cos\theta_{23})+2 q k_3 k_2^2I^{01}(\cos\theta_{23})+2 q k_2 k_3^2I^{10}(\cos\theta_{23}) \right.\nonumber\\
    &+&\left. 
 q^2(k_2^2+k_3^2)I^{00}(\cos\theta_{23})+4q^2 k_2 k_3I^{11}(\cos\theta_{23}) 
 +2 q^3 k_3I^{01}(\cos\theta_{23})\right.\nonumber\\
 &+&\left.2 q^3 k_2I^{10}(\cos\theta_{23})
 +q^4I^{00}(\cos\theta_{23})
    \right],\nonumber\\
    &&
\end{eqnarray}
where  $\cos\theta_{23}=\vec k_2\cdot \vec k_3/k_2k_3$,

\begin{eqnarray}\label{eq:expansionNI}
I^{ij}(\cos\theta_{23})=\sum_{k=0}^\infty \sum_{s=0}^\infty N_{ks} I^{ij}_{ks}(\cos\theta_{23}), 
\end{eqnarray}

\begin{align}
N_{ks}&=\frac{4}{9} \left(\frac{3}{2}+k\right)\left(\frac{3}{2}+s\right)
\left[(r_m q)^k\left(\frac{1}{(r_m q)}\frac{{\rm d}}{{\rm d}(r_m q)} \right)^kW_3(r_m q)\right]\left[(r_m q)^s\left(\frac{1}{(r_m q)}\frac{{\rm d}}{{\rm d}(r_m q)} \right)^sW_3(r_m q)\right]
\nonumber\\
&~~
\cdot \left[(r_m k_2)^k\left(\frac{1}{(r_m k_2)}\frac{{\rm d}}{{\rm d}(r_m k_2)} \right)^kW_3(r_m k_2)\right]\left[(r_m k_3)^s\left(\frac{1}{(r_m k_3)}\frac{{\rm d}}{{\rm d}(r_m k_3)} \right)^s W_3(r_m k_3)\right] ,
\end{align}
and

\begin{eqnarray}
I^{00}_{ks}&=&\left(\frac{1+(-1)^{k+s}}{2}\right)\dfrac{{\rm d}}{{\rm d} \cos\theta_{23}}P_{{\rm min}(k,s)+1}(\cos\theta_{23})\nonumber\\
I^{10}_{ks}&=&
\left(\frac{1-(-1)^{k+s}}{2}\right)\left[\Theta(k-s)\frac{\rm d}{{\rm d} \cos\theta_{23}}P_{k+1}(\cos\theta_{23})+\Theta(s-k) \cos\theta_{23}\frac{\rm d}{{\rm d} \cos\theta_{23}}P_{s+1}(\cos\theta_{23})\right],\nonumber\\
I^{01}_{ks}&=&
\left(\frac{1-(-1)^{k+s}}{2}\right)\left[\Theta(s-k)\frac{\rm d}{{\rm d} \cos\theta_{23}}P_{k+1}(\cos\theta_{23})+\Theta(k-s)\cos\theta_{23} \frac{\rm d}{{\rm d} \cos\theta_{23}}P_{s+1}(\cos\theta_{23})\right],\nonumber\\
I^{11}_{ks}&=&I^{01}_{k+1,s}-(k+2) \left\{\left(\frac{1+(-1)^{k+s}}{2}\right)
\Theta(s-k-1) P_{k+1}(\cos\theta_{23})+\frac{s+1}{2s+3}\delta_{sk}P_{s+1}(\cos\theta_{23})\right\}.\nonumber\\
&&
\end{eqnarray}
This will contribute to an infinite tower of operators to subtract.
The second diagram in Eq. (\ref{two})  gives
\begin{eqnarray}
   I^{(2)}_{\rm two-legs} &=&g_{NL} \frac{108}{25}  \left(\frac{9}{4}\right)^2 P_{\delta}(k_3)P_{\delta}(k_2)\int_{\vec q} P_{\delta}(q)\frac{(|\vec k_1-\vec q|r_m)^2\;W_3(|\vec k_1-\vec q|r_m)}{(qr_m)^2W_3(qr_m)(k_2r_m)^2W_3(k_2r_m)(k_3r_m)^2W_3(k_3r_m)}
   \nonumber\\
    &=&g_{NL} \frac{108}{25}  \left(\frac{16}{81}\right)^2 (k_2r_m)^2 W_3(k_2 r_m)P_{\zeta_g}(k_2) (k_3r_m)^2 W_3(k_3 r_m)P_{\zeta_g}(k_3) \nonumber\\
    &\cdot&
     \int_{\vec q} (qr_m)^2 W_3(q r_m) P_{\zeta_g}(q)\;(|\vec k_1-\vec q|r_m)^2\;W_3(|\vec k_1-\vec q|r_m).
\end{eqnarray}
Using again Eq. (\ref{eq:j_splitting}), we can perform the angular integration, leading to 
\begin{eqnarray}
I^{(2)}_{\rm two-legs} &=& g_{NL} \frac{108}{25}  \left(\frac{16}{81}\right)^2 (k_2r_m)^2 W_3(k_2 r_m)P_{\zeta_g}(k_2) (k_3r_m)^2 W_3(k_3 r_m)P_{\zeta_g}(k_3)\nonumber \\
    &\cdot& \int_{\vec q} (qr_m)^2 W_3^2(q r_m) P_{\zeta_g}(q)\; W_3(k_1r_m)\left[ (qr_m)^2 + (k_1r_m)^2 + \frac{(qr_m)^2}{3} \frac{{\rm d} \ln W_3(k_1r_m)}{{\rm d} \ln k_1 r_m }\right. \nonumber\\
    &+& \left.\frac{(k_1r_m)^2}{3} \frac{{\rm d} \ln W_3(qr_m)}{{\rm d} \ln q r_m } \right].
\end{eqnarray}
If we define the composite operator

\begin{equation}
    \zeta^2(r_m,\vec x_{\rm pk})=
\int\frac{{\rm d}^3 k}{(2\pi)^3}\,
    e^{i\vec k\cdot\vec x_{\rm pk}}W_3(k r_m) (\zeta^2_{g})_{\vec k},
\end{equation}
the operators to be added at this stage are

\begin{equation}\label{eq:twoleg1}
- \frac{9}{25} \cdot \frac{16}{81} \cdot 6\; g_{NL}
     \left( \frac 23 \sigma^2_{\nabla \zeta_g(r_m)} + \frac 16 \frac{{\rm d}\sigma^2_{\nabla \zeta_g(r_m)}}{{\rm d}\ln r_m} \right)r_m^2\nabla^2_{x_{\rm pk}}\zeta^2(r_m,\vec x_{\rm pk})- \frac{9}{25}\cdot 6   \;g_{NL}\; \sigma^2_{\delta(r_m)}\left(1+\frac{1}{3}\frac{{\rm d}}{{\rm d}\ln r_m}\right)\zeta^2(r_m,\vec x_{\rm pk}).
\end{equation}
Finally, from the   third diagram in Eq. (\ref{two})   the operators to add are trivially

\begin{equation}\label{eq:twoleg2}
   \frac{3}{5}f_{NL}\cdot \frac{4}{9}\left[-\frac{12}{5}f_{NL}\cdot \frac{4}{9}\left(\frac{2}{3}\sigma^2_{\nabla\zeta_g(r_m)}+\frac{1}{6}\frac{{\rm d} \sigma^2_{\nabla\zeta_g(r_m)}}{{\rm d}\ln r_m}\right)r_m^2\nabla^2_{x_{\rm pk}}\zeta^2(r_m,\vec x_{\rm pk})+\frac{12}{5}f_{NL}\cdot\frac{5}{3}\cdot\sigma^2_{\delta(r_m)}\zeta^2(r_m,\vec x_{\rm pk})\right].
\end{equation}
Summing up all the contributions from $I^{(2)}_{\rm two-legs}$ and $I^{(3)}_{\rm two-legs}$, and writing only the first terms for $I^{(1)}_{\rm two-legs}$, we obtain at the level of quadratic mixing 
\vskip 0.2cm

\begin{mdframed}
\begin{eqnarray}
\label{secondcomposite}
\left[\delta^2(r_m,\vec x_{\rm pk})\right]_{\text{quadratic }}&=&\delta^2(r_m,\vec x_{\rm pk})-\frac{18}{25}f_{NL}^2 \sigma^2_{\zeta_g(r_m)}\delta^2(r_m,\vec x_{\rm pk})\nonumber\\
&+& \frac{16}{25}f_{NL}^2 \sigma^2_{\nabla\zeta(r_m)}\zeta(r_m,\vec x_{\rm pk})\delta(r_m,\vec x_{\rm pk})\nonumber\\
&+&\frac{256}{675}f_{NL}^2 \sigma^2_{\nabla\zeta(r_m)}\nabla_{x_{\rm pk}^i}\zeta(r_m,\vec x_{\rm pk})\nabla^{x_{\rm pk}^i}\zeta(r_m,\vec x_{\rm pk})\nonumber\\
&-&\left(\frac{32}{75}f_{NL}^2+\frac{64}{225}g_{NL}\right)\left( \frac 23 \sigma^2_{\nabla \zeta_g(r_m)} + \frac 16 \frac{{\rm d}\sigma^2_{\nabla \zeta_g(r_m)}}{{\rm d}\ln r_m} \right)r_m^2\nabla^2_{x_{\rm pk}}\zeta^2(r_m,\vec x_{\rm pk})\nonumber\\
&-&\left[ \frac{54}{25}g_{NL}\; \sigma^2_{\delta(r_m)}\left(1+\frac{1}{3}\frac{{\rm d}}{{\rm d}\ln r_m}\right)-\frac{26}{75}f^2_{NL}\; \sigma^2_{\delta(r_m)}\right]\zeta^2(r_m,\vec x_{\rm pk})\nonumber\\
&+&\cdots.
\end{eqnarray}
\end{mdframed}
Eqs. (\ref{firstcomposite}) and (\ref{secondcomposite}) are the main results of this paper. In the last expression, we have indicated by the dots the infinite series of operators with higher-derivatives applied either to the operators or to the window functions.

To evaluate the impact of the renormalization and operator mixing, one has to recall that, calling $k_\star$ the typical momentum at which PBHs form, one has typically (at the linear level) $k_\star r_m \simeq {\cal O}(3)$~\cite{Germani:2018jgr,Musco:2018rwt}. For instance, for a monochromatic curvature spectrum peaked at $k_\star$, one has   $k_\star r_m\simeq 2.7$. For  a broad spectrum $k_\star$  coincides with the maximum momentum scale, as the PBH mass function peaks at that scale~\cite{DeLuca:2020ioi}, and $k_\star r_m\simeq 3.5$.
One would therefore expect that the higher-derivative terms are suppressed by powers of $1/k_\star r_m$, but having an infinite amount of terms one cannot exclude a priori that the infinite sum of operators give a sizable contribution. 
We will show some numerical insights in the next subsection.

\subsection{The impact of renormalization}
\begin{figure}[t]\centering
\includegraphics[width=0.49 \columnwidth]{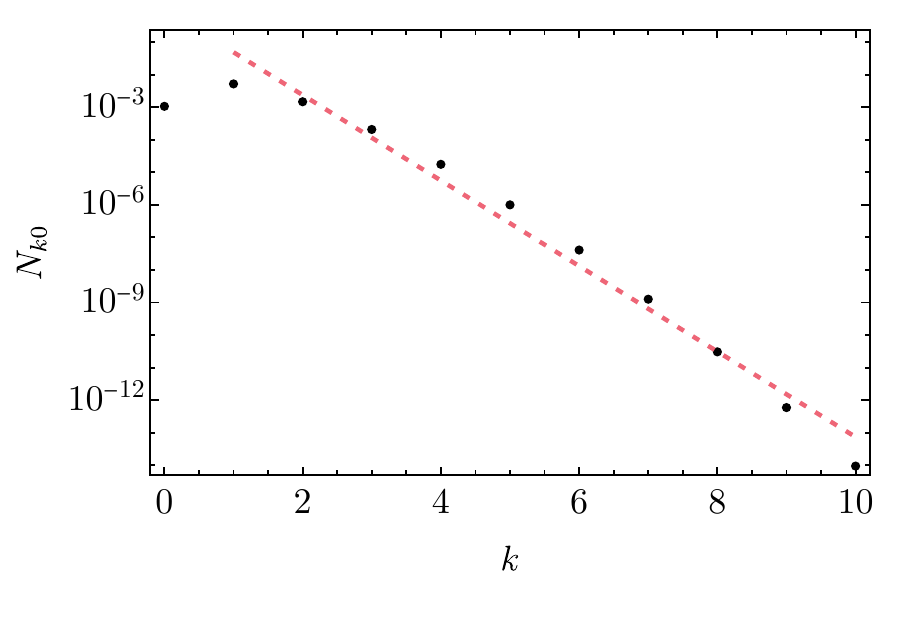}
\includegraphics[width=0.49 \columnwidth]{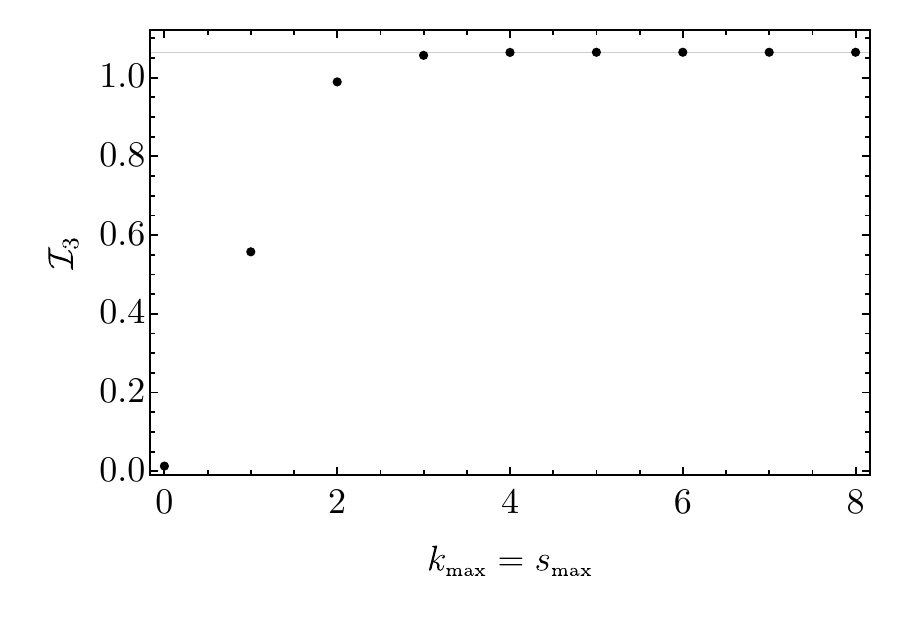}
\vspace{-2em}
\caption{
{\it Left panel:}
Coefficients $N_{ks}$ as a function of k, fixing $s=0$. The red dashed line report a scaling $N_{k0}\propto 1/(k_* r_m)^{3k}$.
{\it Right panel:}
Values of 
${\cal I}_{3}$ as a function of the maximum index included in the sum \eqref{eq:expansionNI}. 
}
  \label{fig:Convergence_test}
\end{figure}

We consider the case of a peaked power spectrum in the curvature perturbation
\be
\frac{k^3}{2\pi^2} P_{\zeta_g}(k)=A_\zeta k_\star\delta(k-k_\star),
\ee
where the amplitude $A_\zeta \simeq 1.8\cdot  10^{-2} $ is fixed by setting a PBH abundance $\beta \simeq 10^{-10}$, neglecting composite operator renormalization.
With this assumption, the integrand in Eq.~\eqref{I3} greatly simplifies as all momenta are fixed to $k_\star$. 
To study the convergence of the result as a function of the sum in $k$ and $s$,  we have plotted $N_{ks}$ in Fig.~\ref{fig:Convergence_test} as a function of index $k$ (and analogous scaling is obtained varying $s$).

The strong hierarchy observed  between the different orders in the sum \eqref{eq:expansionNI} results in a rapidly converging series. We can further investigate this by computing the angular integration 
\begin{equation}
    {\cal I}_3= 4
    \int {\rm d} \theta_{23}
    \left [
    I^{00}(\cos\theta_{23})+
    I^{01}(\cos\theta_{23})+
    I^{10}(\cos\theta_{23})+  
    I^{11}(\cos\theta_{23})
    \right],
\end{equation}
obtained from \eqref{eq:expansionNI} in the limit of narrow power spectrum. We show the result as a function of the maximum index included in the series in Fig.~\ref{fig:Convergence_test}.
We see that already at order $k_\text{\tiny max}=s_\text{\tiny max} = {\cal O} (3)$ the series converges towards the asymptotic value within sufficient accuracy. 

{\bf Correction to the density variance.}
The effect of the renormalization of the composite operator can be quantified as follows.
We first consider its impact on the variance of the field. 
With a narrow power spectrum, the Gaussian component of the curvature perturbation is characterised by a mean peak profile
$\zeta_g(r) = \zeta_{\rm pk} \sin(k_* r)/(k_* r)$. We fix the amplitude of the curvature perturbation in such a way that it corresponds to a realisation of density contrast with the expectation value $\delta_l \approx \sigma_{\delta(r_m)}$.
We find 
\begin{equation}
    \left[\delta^2(r_m,\vec x_{\rm pk})\right]
    = 0.022
    + 0.015 \left ( \frac{f_{NL}}{5/2} \right )
   + 0.019 \left ( \frac{f_{NL}}{5/2} \right )^2
  -0.0016 \left ( \frac{g_{NL}}{25/6} \right ),
\end{equation}
which should be compared with the bare quantity $\delta^2(r_m) \approx \sigma_{\delta(r_m)}^2 = 0.022 $.
This propagates on the smoothed density contrast, which can be estimated as
\begin{equation}
     \delta_m(r_m,\vec x_{\rm pk}) = 
     0.14 
     + 0.00091 \left ( \frac{f_{NL}}{5/2} \right ) 
     - 0.0070 \left ( \frac{f_{NL}}{5/2} \right )^2 
     + 0.00075 \left ( \frac{g_{NL}}{25/6} \right ).
\end{equation}

{\bf Correction to the threshold.}
In order to evaluate the correction to the PBH threshold for collapse, one should estimate the renormalization of the composite operators assuming the fields amplitude reach the threshold $\delta_c$.
We fix the amplitude of the Gaussian curvature perturbation $\zeta_g$ assuming it leads to a collapsing overdensity with threshold amplitude $\delta_m = \delta_c $, where for a narrow spectrum we take $\delta_c = 0.59$ \cite{Musco:2020jjb}.
This corresponds to $\delta_l = 0.88$ and $\zeta_{\rm pk} = 0.68$.
In this case, we find 
\begin{equation}
    \left[\delta^2(r_m,\vec x_{\rm pk})\right]
    = 0.78
    + 0.091 \left ( \frac{f_{NL}}{5/2} \right )
   + 0.66 \left ( \frac{f_{NL}}{5/2} \right )^2
  -0.031 \left ( \frac{g_{NL}}{25/6} \right ),
\end{equation}
while its bare value is $\delta^2(r_m) \approx 0.78 $.
In terms of smoothed density contrast, the renormalisation of the quadratic operators gives
\begin{equation}
     \delta_m(r_m,\vec x_{\rm pk}) = 
    0.59
     + 0.20 \left ( \frac{f_{NL}}{5/2} \right ) 
     - 0.25 \left ( \frac{f_{NL}}{5/2} \right )^2 
     + 0.020 \left ( \frac{g_{NL}}{25/6} \right ).
\end{equation}

\section{Comments and conclusions}
We have shown that in the renormalization of the quadratic power of the smoothed density contrast, a composite operator entering in the calculation of the PBH abundance, leads to the well-known phenomenon of operator mixing due to the non-Gaussian nature of the curvature perturbation. The mixing gives to an  infinite tower of operators
due to the necessary operation of smoothing out  with  the top-hat window function in real space. 
There are some comments we can offer at this stage:

\begin{enumerate}
    \item The calculation of the PBH abundance is not as straightforward as standardly assumed. Not only non-linear corrections entering in the calculation of the PBH from the non-linear radiation transfer function and the determination of the true physical horizon crossing are important and lead 
    to large  uncertainties in the final result \cite{DeLuca:2023tun}, but also the renormalization procedure leads to an infinite tower of operators, making the calculation of the formation probability of the PBH a difficult task. Our findings indicate a large impact on the threshold for PBH formation, depending on the sign of the non-Gaussianity parameters. Since in  ultra-slow inflation the sign of the parameter $f_{NL}$
is always positive \cite{Firouzjahi:2023xke}, our results indicate that the threshold  might in fact decrease due to quadratic corrections ${\cal O}(f_{NL}^2)$, thus increasing the PBH  abundance and reversing the ${\cal O}(f_{NL})$ impact.

    \item The one-loop operator mixing leads to an infinite tower of operators, made of quadratic powers of the density contrast $\delta$ and higher-derivatives thereof. Furthermore, the mixing occurs as well with non-local operators if considered from the point of view of the smoothed density contrast (or local if expressed in terms of the original curvature perturbation). 
    Maybe one can intuitively understand why an infinite tower of operators is needed by the following argument. Around the  peak of the curvature perturbation, one can perform  a rotation of the coordinate axes to be aligned with the principal axes of length
$\lambda_i$ ($i=1,2,3$) of the constant-curvature perturbation ellipsoid and Taylor expanding up to second-order gives~\cite{Bardeen:1985tr} 
\be
\zeta(r)\simeq \zeta_{\rm pk}-\frac{1}{2}\sum_{i=1}^3\lambda_i\left(x_i-x_i^{\rm pk}\right)^2, 
\ee
and
\be
\delta(r_m,\vec x_{\rm pk})= -\frac{4}{3}r_m\zeta'(r_m)\simeq \frac{8}{3}\left[\zeta_{\rm pk}-\zeta(r_m)\right].
\ee
This  shows that  already at the linear level, the statistics of the 
smoothed density contrast calculated in a volume of radius $r_m$ demands knowing  all  correlations of the curvature perturbation in two different spatial points \cite{DeLuca:2022rfz} and therefore all its  gradients.

\item In an idealized scenario, one might envisage to calculate the PBH abundance by involving only superhorizon physics to avoid the complications arising at horizon crossing, and thus relying on the fact that the comoving number of peaks which eventually will collapse into PBHs at horizon crossing remains constant on superhorizon scales. The   renormalization  procedure however must be applied  as well on superhorizon scales as composite operators probe short scales and  the latter may combine  in the loops  to generate long modes.  Therefore, even on superhorizon scales calculating the PBH formation probability  might be more involved than naively thought.

    \item The probability of forming a PBH should not depend on any smoothing procedure, and therefore should be independent of any window function we decide to force into the calculation of the PBH abundance. The renormalization procedure is exactly performed to leave no sign of such arbitrary choice. Being the non-linear density contrast $\delta_\com(\vec x,t)$  on  comoving orthogonal slicings (\ref{deltaNL}) a composite operator in terms of the time independent curvature perturbation $\zeta(\vec x)$, one concludes that its renormalization should be performed in order to lead to a result independent of the choice of the smoothing procedure.

\end{enumerate}
We intend to elaborate about these points in future work.

\vskip 0.5cm
\noindent
\centerline{{\bf Acknowledgments.}} 
\vskip 0.3cm
\noindent
We thank V. De Luca for useful discussions. A.I. and A.R.  acknowledge support from the  Swiss National Science Foundation (project number CRSII5\_213497).
D. P. and A.R. are supported by the Boninchi Foundation for the project ``PBHs in the Era of GW Astronomy''.

\newpage
\appendix
\section{Integrals involving the Gegenbauer polynomials}
\subsection{Integral of one window function}
The  window function $W_3(|\vec k-\vec q|r_m)$, defined as
\begin{equation}
    W_3(x) = 2^{3/2}\Gamma(5/2) \frac{J_{3/2}(x)}{x^{3/2}},
\end{equation}
can be treated in the following way. First we use the property

\begin{equation}
    \frac{J_\nu(|\vec l_1+\vec l_2|)}{|\vec l_1+\vec l_2|^\nu}=2^\nu\Gamma(\nu)\sum_{k=0}^\infty(\nu + k)\frac{J_{\nu+k}(l_1)}{l_1^{\nu}}\frac{J_{\nu+k}(l_2)}{l_2^{\nu}}C_k^\nu(\cos\theta),
\end{equation}
where $C_k^\nu(\cos\theta)$ are the Gegenbauer polynomials and $\theta$ is the angle between the 
vectors $\vec l_1$ and $\vec l_2$.  In general
\begin{equation}
    \int_0^{\pi} {\rm d}\theta \,C^{\nu}_k (\cos \theta) \sin^{2\nu}\theta  = 0,\quad \forall \;k \neq 0.
\end{equation}
Therefore, 
\begin{eqnarray}
    \int \frac{{\rm d}\Omega}{4\pi} \frac{J_{1/2}(|\vec l_1 + \vec l_2|)}{|\vec l_1 + \vec l_2|^{1/2}} &=&   2^{1/2}\Gamma(1/2)   \frac 12 \frac{J_{1/2}(l_1)}{l_1^{1/2}}\frac{J_{1/2}(l_2)}{l_2^{1/2}} \int \frac{{\rm d}\Omega}{4\pi} C_0^{1/2}(\cos\theta) \;+\nonumber\\
    &+&2^{1/2}\Gamma(1/2) \sum_{k=1}^{\infty} (1/2 + k) \frac{J_{1/2+k}(l_1)}{l_1^{1/2}}\frac{J_{1/2+k}(l_2)}{l_2^{1/2}}  \int \frac{{\rm d}\Omega}{4\pi} C_k^{1/2}(\cos\theta).
\end{eqnarray}
Only the first term in the sum is not zero, leading to the simple relation
\begin{equation}
\label{eq:j1/2}
    \int \frac{{\rm d}\Omega}{4\pi} \frac{J_{1/2}(|\vec l_1 + \vec l_2|)}{|\vec l_1 + \vec l_2|^{1/2}} = 2^{1/2}\Gamma(1/2)   \frac 12 \frac{J_{1/2}(l_1)}{l_1^{1/2}}\frac{J_{1/2}(l_2)}{l_2^{1/2}}.
\end{equation}
Now we can use the general relations
\begin{equation}
    l\frac{{\rm d}}{{\rm d} l} \frac{J_{D/2 -1} (l)}{l^{D/2 -1 }}= - \frac{J_{D/2}(l)}{l^{D/2-2}},
\end{equation}
\begin{equation}
\label{fu}
     \frac{J_{D/2 -1} (l)}{l^{D/2 -1 }}=  l\frac{{\rm d}}{{\rm d} l}\frac{J_{D/2}(l)}{l^{D/2}} + D \frac{J_{D/2}(l)}{l^{D/2}},
\end{equation}
to transform the equation above from $J_{1/2}$ to $J_{3/2}$, using the relation

\be
l_1\partial_{l_1}\frac{J_{1/2}(|\vec l_1+\vec l_2|)}{|\vec l_1+\vec l_2|^{1/2}}=-(l_1^2+\vec l_1\cdot \vec l_2)\frac{J_{3/2}(|\vec l_1+\vec l_2|)}{|\vec l_1+\vec l_2|^{3/2}}.
\ee
We take  the derivative with respect to $l_1$ of Eq. (\ref{eq:j1/2}), leading to 

\begin{equation}
    \int \frac{{\rm d} \Omega}{4 \pi} W_3(|\vec l_1 - \vec l_2|)\left( 1- \frac{\vec l_1 \cdot \vec l_2}{l_1^2}\right)= W_3(l_1) W_3(l_2) + W_3(l_1) \; \frac{l_2}{3}\frac{{\rm d}}{{\rm d} l_2} W_3(l_2).
\end{equation}
Multiplying both sides with $l_1^2$ and summing the same relation exchanging $l_1$ with $l_2$, we get 
\begin{align}
\label{eq:j_splitting}
    &\int \frac{{\rm d} \Omega}{4 \pi} W_3(|\vec l_1 - \vec l_2|)\left( l_1^2 + l_2^2- 2\vec l_1 \cdot \vec l_2\right)=\int \frac{{\rm d} \Omega}{4 \pi} W_3(|\vec l_1 - \vec l_2|)|\vec l_1 - \vec l_2|^2= \nonumber\\
    &=W_3(l_1) W_3(l_2)(l_1^2 + l_2^2) + W_3(l_1) \; \frac{l_1^2 l_2}{3}\frac{{\rm d}}{{\rm d} l_2} W_3(l_2) + W_3(l_2) \; \frac{l_2^2 l_1}{3}\frac{{\rm d}}{{\rm d} l_1} W_3(l_1)=\nonumber\\
    &= W_3(l_1) W_3(l_2) \left( l_1^2 + l_2^2 +  \frac{l_1^2 l_2}{3W_3(l_2)}\frac{{\rm d}}{{\rm d} l_2} W_3(l_2) + \; \frac{l_2^2 l_1}{3W_3(l_1)}\frac{{\rm d}}{{\rm d} l_1} W_3(l_1)\right).
\end{align}

\subsection{Integral with two window functions}
\label{appendix:2window_fun}
For the  composite operator with two external legs, we have to evaluate the following momentum integral 

\begin{align}
    I&=\int_{\vec {l}_1} |\vec l_1+\vec l_2|^2W_3(|\vec l_1+\vec l_2|) |\vec l_1+\vec l_3|^2W_3(|\vec l_1+\vec l_3|)\nonumber\\
    &=\int_{\vec {l}_1} ( l_1^2+ l_2^2+2 l_1 l_2\cos\theta_{12})W_3(|\vec l_1+\vec l_2|) ( l_1^2+ l_3^2+2l_1 l_3\cos\theta_{13})W_3(|\vec l_1+\vec l_3|)\nonumber\\
    &=\int_{\vec {l}_1}\Big(A_{00}+A_{10} \cos\theta_{12}+A_{01}\cos\theta_{13}+A_{11}\cos\theta_{12}\cos\theta_{13}\Big)
W_3(|\vec l_1+\vec l_2|)W_3(|\vec l_1+\vec l_3|)\nonumber \\
&= 2^3\Gamma^2(5/2)\int_{\vec {l}_1}\Big(A_{00}+A_{10} \cos\theta_{12}+A_{01}\cos\theta_{13}+A_{11}\cos\theta_{12}\cos\theta_{13}\Big) \frac{J_{3/2}(|\vec l_1+\vec l_2|)}{|\vec l_1+\vec l_2|^{3/2}} \frac{J_{3/2}(|\vec l_1+\vec l_3|)}{|\vec l_1+\vec l_3|^{3/2}},
&
\label{I0}
\end{align}
where 
\begin{eqnarray}
&&A_{00}=l_1^4 +l_1^2(l_2^2+l_3^2)+l_2^2l_3^2,~~~~~A_{10}=2l_1l_2(l_1^2+l_3^2),\nonumber\\
&&A_{01}=2l_1l_3(l_1^2+l_2^2), ~~~~~~~~~~~~~~~~~A_{11}=4l_1^2l_2l_3.
\end{eqnarray}
We have parametrized the vectors $\vec{l_i}$ as 
\begin{align}
    &\vec l_2 = l_2\; (0,\; 0,\;1),\nonumber\\
    &\vec l_3 = l_3\; (0,\; \sqrt{1-y^2},\;y),\nonumber\\
    &\vec l_1 = l_1\; (\cos\beta\sqrt{1-x^2}, \;\sin\beta\sqrt{1-x^2}, \;x),
\end{align}

so that \begin{equation}
\label{theta}
    \cos \theta_{12} = x,~~~~~ \cos \theta_{13} = xy + \sqrt{1-y^2}\sqrt{1-x^2} \sin \beta.
\end{equation}
Using the addition theorem
\begin{align}
\label{expansion}
    \frac{J_\nu(|\vec l_1+\vec l_2|)}{|\vec l_1+\vec l_2|^\nu} \frac{J_\nu(|\vec l_1+\vec l_3|)}{|\vec l_1+\vec l_3|^\nu} &=(2^\nu\Gamma(\nu))^2 \sum_{k=0}^\infty \sum_{s=0}^\infty \frac{J_{\nu+k}(l_1)}{l_1^{\nu}}\frac{J_{\nu+k}(l_2)}{l_2^{\nu}}\frac{J_{\nu+s}(l_1)}{l_1^{\nu}}\frac{J_{\nu+s}(l_3)}{l_3^{\nu}} \nonumber \\
    &\cdot(\nu+k)(\nu+s)C_k^\nu(\cos\theta_{12}) C_s^\nu(\cos\theta_{13}),
\end{align}
we can express the integral (\ref{I0}) as 
\begin{align}
I&=\int_{\vec {l}_1}\Big(A_{00}+A_{10} \cos\theta_{12}+A_{01}\cos\theta_{13}+A_{11}\cos\theta_{12}\cos\theta_{13}\Big) \sum_{k=0}^\infty \sum_{s=0}^\infty 9\pi^2 \left(\frac{3}{2}+k\right)
\nonumber \\
&\cdot\left(\frac{3}{2}+s\right) \frac{J_{3/2+k}(l_1)}{l_1^{3/2}}\frac{J_{3/2+k}(l_2)}{l_2^{3/2}}\frac{J_{3/2+s}(l_1)}{l_1^{3/2}}\frac{J_{3/2+s}(l_3)}{l_3^{3/2}} C_k^{3/2}(\cos\theta_{12}) C_s^{3/2}(\cos\theta_{13}).
\end{align}
Therefore, the integral (\ref{I0}) can be written as 
\begin{eqnarray}
I=\sum_{i,j=0}^1\int \frac{l^2{\rm d}l}{2\pi^2}A_{ij}I^{ij},
\end{eqnarray}
where 
\begin{eqnarray}
I^{ij}=\sum_{k=0}^\infty \sum_{s=0}^\infty N_{ks} I^{ij}_{ks}(y), 
\end{eqnarray}

\begin{align}
N_{ks}&=9\pi^2 \left(\frac{3}{2}+k\right)\left(\frac{3}{2}+s\right) \frac{J_{3/2+k}(l_1)}{l_1^{3/2}}\frac{J_{3/2+k}(l_2)}{l_2^{3/2}}\frac{J_{3/2+s}(l_1)}{l_1^{3/2}}\frac{J_{3/2+s}(l_3)}{l_3^{3/2}}\nonumber \\
&=\frac{4}{9} \left(\frac{3}{2}+k\right)\left(\frac{3}{2}+s\right)
\left[l_1^k\left(\frac{1}{l_1}\frac{{\rm d}}{{\rm d}l_1} \right)^kW_3(l_1)\right]\left[l_1^s\left(\frac{1}{l_1}\frac{{\rm d}}{{\rm d}l_1} \right)^sW_3(l_1)\right]
\nonumber\\
&~~
\times \left[l_2^k\left(\frac{1}{l_2}\frac{{\rm d}}{{\rm d}l_2} \right)^kW_3(l_2)\right]\left[l_3^s\left(\frac{1}{l_3}\frac{{\rm d}}{{\rm d}l_3} \right)^s W_3(l_3)\right] ,
\end{align}
and 
\begin{eqnarray}
I^{ij}_{ks}(y=\cos\theta_{23})=\int\frac{{\rm d}\Omega}{4\pi}(\cos\theta_{12})^i(\cos\theta_{13})^j C_k^{3/2}(\cos\theta_{12}) C_s^{3/2}(\cos\theta_{13}).
\end{eqnarray}
Therefore, in order to calculate Eq. (\ref{I0}), we need to calculate the four integrals
\begin{eqnarray}
I^{00}_{ks}(y), ~I^{10}_{ks}(y), ~ I^{01}_{ks}(y), ~~~\mbox{and}~~~I^{11}_{ks}(y),
\end{eqnarray}
which we will do in the following. 
Inserting the angles (\ref{theta}) and using the fact that
\begin{align}
\int_0^{2\pi}\frac{{\rm d}\beta}{2\pi}\, C_k^{3/2}\left(xy + \sqrt{1-y^2}\sqrt{1-x^2} \sin \beta\right) &=
  \left\{\begin{array}{lll}
   {\displaystyle{\sum_{n={\rm even}}^k}}(2n+1)P_{n}(x) P_{n}(y)&~{\rm if}&~ k=\mbox{even},\\
   &&\\
   {\displaystyle{\sum_{n={\rm odd}}^k}}(2n+1)P_{n}(x) P_{n}(y)&~{\rm if}&~ k=\mbox{odd},\\
  \end{array}\right.  \label{db}
\end{align}
and 
\begin{align}
C_k^{3/2}(x)&=\frac{{\rm d}}{{\rm d} x}P_{k+1}(x)\nonumber \\&=(2k+1)P_k(x)+\Big(2(k-2)+1\Big)P_{k-2}(x)+\Big(2(k-4)+1\Big) P_{k-4}(x)+\cdots, \label{ck}
\end{align}
the integration over the angles gives
\begin{align}
 I^{00}_{ks}&= \int\frac{{\rm d}\Omega}{4\pi} C_k^{3/2}(x) C_s^{3/2}\left(xy + \sqrt{1-y^2}\sqrt{1-x^2} \sin \beta\right)= 
 \nonumber\\ &\hspace{4cm}=
  \left\{\begin{array}{lll}
   \dfrac{{\rm d}}{{\rm d} y}P_{{\rm min}(k,s)+1}(y)&{\rm if}& (k,s)\in  2{\mathbb N},\\&&\\
   \dfrac{{\rm d}}{{\rm d} y}P_{{\rm min}(k,s)+1}(y)&{\rm if}& (k,s)\in  2{\mathbb N}+1,\\
   &&\\
   0 &{\rm otherwise}, &\\
  \end{array}\right. ,
  \end{align}
or
\begin{eqnarray}
I^{00}_{ks}=\left(\frac{1+(-1)^{k+s}}{2}\right)\dfrac{{\rm d}}{{\rm d} y}P_{{\rm min}(k,s)+1}(y).
\end{eqnarray}
The next integral to evaluate is
\begin{equation}
    I_{ks}^{10}(y)=\int\frac{{\rm d}\Omega}{4\pi} \,x\, C_k^{3/2}(x) C_s^{3/2}(\cos\theta_{13}) .
\end{equation}
To proceed, we need the recursive relation of Legendre polynomials 
\begin{eqnarray}
x\frac{{\rm d} P_{k+1}(x)}{{\rm d} x}=\frac{{\rm d} P_{k+2}(x)}{{\rm d} x}-(k+2)P_{k+1}(x),
\end{eqnarray}
which can be written as 
\begin{eqnarray}
x\,C_k^{3/2}(x)=C^{3/2}_{k+1}-(k+2)P_{k+1}(x).  \label{s1}
\end{eqnarray}
Then, using Eq. (\ref{db}), we can easily see that  
\begin{eqnarray}
\int_0^{2\pi}\frac{{\rm d}\Omega}{4\pi}\,P_{k+1}(x) 
C_s^{3/2}(\cos\theta_{13})=\left\{\begin{array}{lll}
   \Theta(s-k-1)\, P_{k+1}(y)&{\rm if}& (k,s)\in  2{\mathbb N},\\&&\\
   \Theta(s-k-1)\, P_{k+1}(y)&{\rm if}& (k,s)\in  2{\mathbb N}+1,\\
   &&\\
   0 &{\rm otherwise}, &\\
  \end{array}\right. \label{s2}
\end{eqnarray}
where $\Theta(n)=0$ for $n<0$ and $\Theta(n)=1$ for $n\geq 0$. 
Therefore, using Eqs. (\ref{s1}) and (\ref{s2}), we find that 
\begin{eqnarray}
I^{10}_{ks}=
\left(\frac{1-(-1)^{k+s}}{2}\right)\left[\Theta(k-s)\frac{\rm d}{{\rm d} y}P_{k+1}(y)+\Theta(s-k)y \frac{\rm d}{{\rm d} y}P_{s+1}(y)\right] 
\end{eqnarray}
To calculate $I^{01}$ we need the relation 
\begin{align}
\int_0^{2\pi}\frac{{\rm d}\beta}{2\pi}\,& \cos\theta_{13}C_k^{3/2}\left(\cos\theta_{13}\right)=\nonumber\\
&  =
  \left\{\begin{array}{lll}
  2(k+1)P_{k+1}(x)P_{k+1}(y)+
   {\displaystyle{\sum_{n={\rm odd}}^k}}2(2n+1)P_{n}(x) P_{n}(y)&~{\rm if}&~ k=\mbox{even},\\
   &&\\
   2(k+1)P_{k+1}(x)P_{k+1}(y)+
   {\displaystyle{\sum_{n={\rm even}}^k}}2(2n+1)P_{n}(x) P_{n}(y)&~{\rm if}&~ k=\mbox{odd},\\
  \end{array}\right.  \label{db1}
\end{align}
Then, by using Eq. (\ref{ck}), we find that 
the integral 
\begin{eqnarray}
I^{01}_{ks}&=&\int\frac{{\rm d}\Omega}{4\pi}\,C^{3/2}_k(x)\cos\theta_{13} 
C_s^{3/2}(\cos\theta_{13})
\end{eqnarray}
turns out to be 
\begin{equation}
I^{01}_{ks}=
\left(\frac{1-(-1)^{k+s}}{2}\right)\left[\Theta(s-k)\frac{\rm d}{{\rm d} y}P_{k+1}(y)+\Theta(k-s)y \frac{\rm d}{{\rm d} y}P_{s+1}(y)\right].
 \label{s4}
\end{equation}
The final integral $I^{11}_{ks}$ can be calculated using Eq. (\ref{s1}). The result is
\begin{eqnarray}
I^{11}_{ks}=I^{01}_{k+1,s}-(k+2) \left\{\left(\frac{1+(-1)^{k+s}}{2}\right)
\Theta(s-k-1) P_{k+1}(y)+\frac{s+1}{2s+3}\delta_{sk}P_{s+1}(y)\right\}.
\end{eqnarray}

\section{Non-local operators}
\label{AppendixB}
To derive the expression (\ref{firstcomposite}), we start Eq. (\ref{start}) which can 
 be rewritten as
\begin{equation}
    \frac{\delta(r_m,\vec x_{\rm pk})}{r_m^2} = 
    \frac 49 \nabla^2_{x_{\rm pk}}\int_{\vec k} \; e^{i \vec k\cdot \vec x_{\rm pk}}  W_3(k r_m)\zeta_{\vec k}.
\end{equation}
We now take the  derivative $r_m\partial_{r_m}$ of both sides to get
\begin{equation}
    \frac{1}{r_m}\partial_{r}\delta(r,\vec x_{\rm pk})\Big|_{r=r_m} = -2\frac{\delta(r_m,\vec x_{\rm pk})}{r_m^2} + \frac{1}{r_m}  \partial_{r} \delta(r,\vec x_{\rm pk})\Big|_{r=r_m}=
    \frac 49 \nabla^2_{x_{\rm pk}}\int_{\vec k} \; e^{i \vec k\cdot \vec x_{\rm pk}} \zeta_{\vec k}\, r_m\partial_{r} W_3(k r)\Big|_{r=r_m} .
\end{equation}
By using Eq. (\ref{max}) we get 

\begin{equation}
    \partial_{r} \delta(r,\vec x_{\rm pk})\Big|_{r=r_m}=0,
\end{equation}
and consequently
\begin{equation}
    -2\frac{\delta(r_m,\vec x_{\rm pk})}{r_m^2} =
    \frac 49 \nabla^2_{x_{\rm pk}}\int_{\vec k} \; e^{i \vec k\cdot \vec x_{\rm pk}}  \zeta_{\vec k}\,r_m\partial_{r} W_3(k r)\Big|_{r=r_m}
\end{equation}
or 
\begin{equation}
    r_m\partial_{r} W_3(k r)\Big|_{r=r_m} \zeta_{\vec k}=-\frac{9}{2r_m^2} \int_{\vec x_{\rm pk}} \; e^{-i \vec k\cdot \vec x_{\rm pk}}\; \nabla^{-2}_{x_{\rm pk}}\delta(r_m,\vec x_{\rm pk})
\end{equation}
which is a non-local operator.

\newpage

\bibliographystyle{JHEP}
\bibliography{main}

\end{document}